\documentclass[sigconf]{acmart}

\setcopyright{none}
\renewcommand\footnotetextcopyrightpermission[1]{}

\AtBeginDocument{%
  \providecommand\BibTeX{{%
    \normalfont B\kern-0.5em{\scshape i\kern-0.25em b}\kern-0.8em\TeX}}}

\setcopyright{acmcopyright}
\copyrightyear{2024}
\acmYear{2024}
\acmDOI{XXXXXXX.XXXXXXX}

\acmConference[DAC ’24]{Make sure to enter the correct
  conference title from your rights confirmation emai}{June 23--27,
  2024}{San Francisco, CA, USA}
\acmPrice{15.00}
\acmISBN{978-1-4503-XXXX-X/18/06}

\usepackage{subfigure}
\usepackage{caption}
\begin{document}

\title{\fontsize{16}{24}\selectfont HALO-CAT: A \underline{H}idden Network Processor with \underline{A}ctivation-\underline{LO}calized \underline{C}IM \underline{A}rchitecture and Layer-Penetrative \underline{T}iling}

\author{Yung-Chin Chen}
\affiliation{%
  \institution{National Taiwan University/Keio University}
  \country{Taiwan/Japan}
  }
\email{jim.chen.work@gmail.com}

\author{Shimpei Ando}
\affiliation{%
  \institution{Keio University}
  \country{Japan}
  }
\email{shimpeiando@keio.jp}

\author{Daichi Fujiki}
\affiliation{%
  \institution{Keio University}
  \country{Japan}
  }
\email{dfujiki@keio.jp}

\author{Shinya Takamaeda-Yamazaki}
\affiliation{%
  \institution{The University of Tokyo}
  \country{Japan}
  }
\email{shinya@is.s.u-tokyo.ac.jp}

\author{Kentaro Yoshioka}
\affiliation{%
  \institution{Keio University}
  \country{Japan}
  }
\email{kyoshioka47@keio.jp}

\vspace{-2mm}
\begin{abstract}
In addressing the 'memory wall' problem in neural network hardware acceleration, Hidden Networks (HNN) emerge as a promising solution by generating weights directly on-chip, thus significantly reducing DRAM access. Nevertheless, existing HNN processors face challenges with energy efficiency due to large on-chip storage needs and redundant activation movement. 

To tackle these issues, we developed HALO-CAT, a software-hardware co-designed processor optimized for HNNs. 
On the algorithmic front, we present Layer-Penetrative Tiling (LPT) to adapt to extreme cross-layer computation and tiling to substantially reduce the size of on-chip storage.
On the architecture side, we propose an activation-localized (AL) computing-in-memory processor that maximally preserves the locality of activations, thereby minimizing data movement.
HALO-CAT achieves a 14.2$\times$ reduction in activation on-chip memory, a 1.6$\times$ cut down in the number of activation accesses, and an estimated 17.8$\times$ reduction in energy consumption attributed to data communication, compared to the baseline design. Moreover, we maintain an accuracy loss of merely 1.5\% in Resnet50@ImageNet.

\end{abstract}


\maketitle
\pagestyle{plain}

\section{Introduction}
Neural networks (NN) have demonstrated exceptional capabilities across various domains. However, their true potential is still underexplored, especially in energy-constrained applications. This limitation is primarily due to significant data movement inherent in NN operation, which leads to the 'memory wall' problem. Overcoming this challenge is crucial upon achieving energy-efficient NN processing.
The disparity in computing and communication energy is significant: as per \cite{Interstellar}, a 16-bit single Multiply-and-Accumulate (MAC) operation consumes only 0.075pJ, whereas a DRAM access costs over 2000 times more energy, requiring 200pJ under 28nm CMOS process. The high energy associated with I/O communication for NN parameters significantly constrains the system-level efficiency. This limitation negates potential benefits of on-chip data reuse and enhanced computing unit capabilities. 
This issue is exemplified in standard NN accelerators like \cite{DianNao}, where DRAM access alone accounts for more than 95\% of the total energy use. 

The primary challenge in NN processing lies in the distinct mismatch between NN weight sizes and the memory capacity of ASICs. In area-constrained scenarios, the accelerator on-chip SRAM is limited to several 100 kBs. This capacity is substantially smaller compared to the size of typical NNs, which can easily exceed the ASIC's handling capacity by orders of magnitude. For instance, ResNet18 is composed of 11.4M weight parameters, and ResNet50 has even more, with 23.9M. Therefore, typical NN accelerators load these weight parameters from the DRAM repetitively, significantly constraining the system-level efficiency.  
%
Even in accelerators designed to minimize external memory access \cite{SmartExchange}, incorporating software-oriented strategies such as sparsification and weight decomposition, a substantial 50 to 85\% of energy is still expended on off-chip access, which significantly hampers the system-level power efficiency.
This highlights the critical need for further advancements in software-hardware co-design to more effectively confront and overcome the persistent 'memory wall' challenge in current computing architectures.

\begin{figure*}[tbp]
\centerline{\includegraphics[width=18cm]{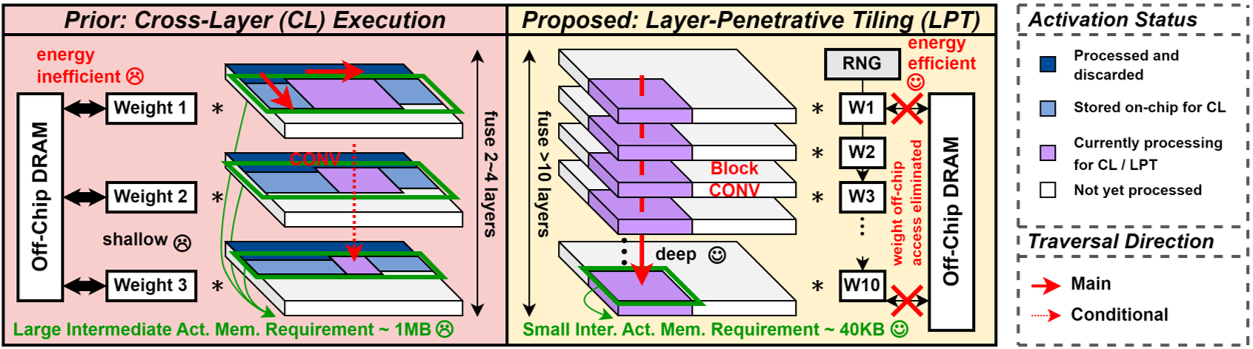}}
\captionsetup{font=footnotesize}
\caption{Comparison between CL execution and our proposed LPT. Leveraging the advantages of HNN on-chip weight generation and blocked HNN, off-chip weight access and activation data dependency can be eliminated. LPT achieves efficient penetration through more than 10 layers. The reinforced tiling approach minimizes the size of intermediate activation from 1MB to 40KB, leading to a 26$\times$ reduction.
}
\label{algorithm_concept}
\end{figure*}


The Hidden Network (HNN) \cite{HNN} represents a breakthrough in software-hardware co-design, targeting the challenges posed by the 'memory wall'. At the core of HNN's methodology is the use of a sparse binary supermask over a randomly weighted network, which selectively retains vital weights while discarding non-essential ones. Remarkably, the network's reliance on entirely random weights enables the generation of significant weight parameters directly on-chip using standard random number generators (RNGs). This principle significantly reduces the need for energy-intensive off-chip weight access, thereby enhancing system-level efficiency.

The initial evaluation of introducing HNNs for accelerator design has been validated in Hiddenite \cite{Hiddenite}, demonstrating reduction in off-chip data access. Nevertheless, the Hiddenite framework's energy efficiency is compromised due to several critical limitations in its algorithmic approach and architecture design. 
\textbf{Algorithmically}, Hiddenite follows traditional CNN structures, leading to considerable overhead in maintaining intermediate results on-chip. For instance, managing ResNet50 operations on-chip requires 1MB of SRAM for activation storage, consuming over 80\% of the chip's core area and significantly impacting power. 
\textbf{Architecturally}, Hiddenite has excessive activation data movement due to the suboptimal reuse of output data, resulting in redundant storage and retrieval between the computing unit and memory. This emphasizes the need for an activation-centric architecture to reduce the data footprint.

To optimize the performance of HNN in hardware acceleration, HALO-CAT introduces a targeted software-hardware co-design. This approach strategically combines algorithmic and architectural advancements to effectively address the unique challenges and harness the strengths of HNNs in a hardware setting.
 \\\textbf{(1) Algorithm Realm:} We propose \textit{Layer-Penetrative Tiling (LPT)} which integrates extensive tiling and cross-layer (CL) computing into the HNN framework. By doing so, LPT effectively reduces the size of intermediate data, which is a key in optimizing on-chip storage demands and improving overall efficiency.
 \\\textbf{(2) Architecture Realm:} We propose an \textit{activation-localized (AL)} compute-in-memory (CIM) processor to maximize the HNN inference efficiency. The AL dataflow is designed specifically for HNN operations and ensures that activations are kept local. By bridging the output features to the subsequent layer, we minimize the corresponding activation movement. 

\vspace{-2mm}
\section{Backgrounds}
\subsection{Cross-Layer (CL) Execution} 
Hardware acceleration of Neural Networks (NNs) typically follows a layer-by-layer processing approach, where each layer's computation starts only after completing the previous one. This method, while straightforward, leads to large storage needs for intermediate results and increased latency.
An alternative approach under exploration is cross-layer (CL) computing, also referred to as depth-first (DF) execution \cite{DepthFirst, DepFiN} or layer-fusion \cite{LayerFusion, diana}. This approach initiates the computation of the next layer's output pixel as soon as the necessary input pixels are available, rather than waiting for the entire feature map's completion. A similar commonly-used technique is tiling \cite{LayerFusion, DeFiNES}, which, in the context of CL computing, involves blockwise traversal of the feature map to reduce the maximum data size required at any given time, thereby diminishing storage requirements\cite{DepthFirst, DepFiN}.

However, the limitations on achievable layer depth in existing CL computing are evident, as presented in Fig. \ref{algorithm_concept}. Despite being labeled as 'depth-first,' the depth it can explore is somewhat shallow, particularly in area-constrained environments. Normally, fusing more than three CONV3x3 layers is generally inefficient, as seen in \cite{Hiddenite} and \cite{diana}, where the former involves only one CONV3x3 layer and the latter fuses four, but with significant overhead in activation memory access.
This is primarily attributed to the following two reasons: 
\\
\textbf{Data Dependency Issue}:
In Cross-Layer (CL) computing, the reliance on data from adjacent tiles for convolution operations increases as the depth of fused layers grows. This results in an escalation of on-chip memory demands and reduced efficiency \cite{BlockConv}.
\\
\textbf{Weight Reloading Issue}:
Multiple fused layers in CL computing lead to an increase in weight parameters. With limited weight buffer sizes in standard ASICs, this necessitates frequent reloading of weights for different input portions, increasing energy consumption and limiting the feasible depth of fused layers.

Inspired by the spirit of CL computing, we propose \textit{Layer-Penetrative Tiling (LPT)}, an approach uniquely enabled by the attributes of HNNs. LPT represents a profound implementation of CL tiling, demonstrating the capability to efficiently process through more than 10 layers while significantly reducing the need for on-chip storage.  By leveraging the inherent structure of HNNs \cite{HNN} and incorporating block convolution \cite{BlockConv}, LPT effectively overcomes the challenges traditionally associated with layer depth and storage requirements.

\subsection{Hidden Network (HNN)}
\begin{figure}[tbp]%
\centering
\subfigure[]{%
\label{HNN_intro}%
\includegraphics[height=3.7cm]{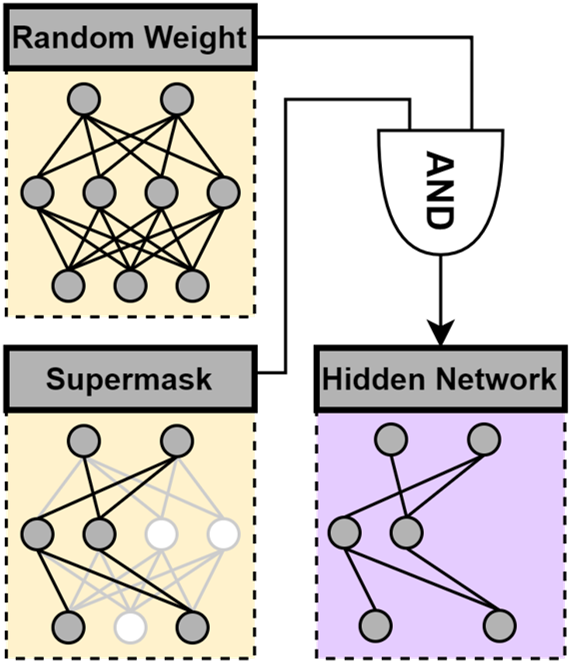}}
\
\subfigure[]{%
\label{blockconv_intro}%
\includegraphics[height=3.7cm]{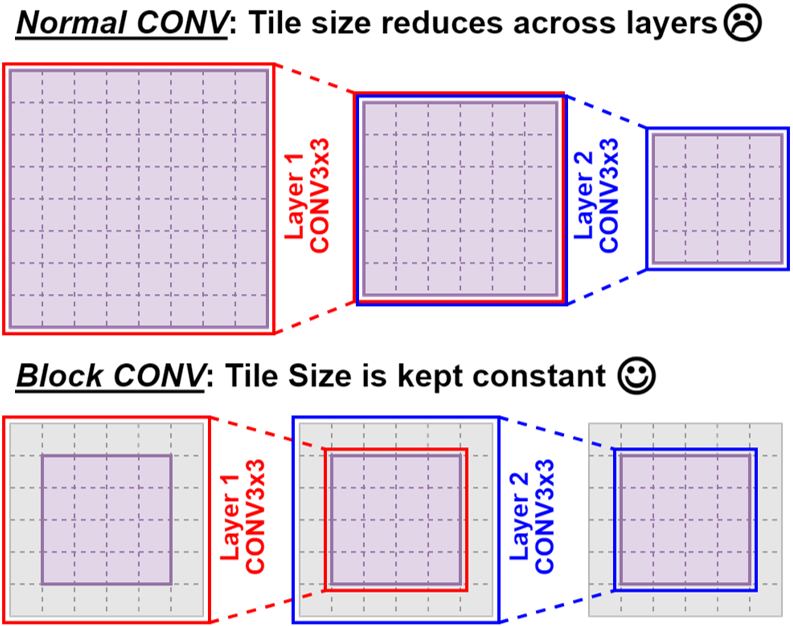}}%
\vspace{-4mm}
\captionsetup{font=footnotesize}
\caption{(a) The fundamentals of HNN. (b) The algorithm of block convolution. The tile size is kept constant with fused layers via inner-tile zero-padding, eliminating the need for data from adjacent tiles.}
\vspace{-4mm}
\end{figure}

While the Lottery Ticket Hypothesis \cite{LotteryTicketHypothesis} inspired researchers to bravely seek valuable parameters from the start to reduce training costs, Hidden Network (HNN) \cite{HNN} takes an even bolder approach by directly exploring a subnetwork on an untrained, or random-weighted network.
The development of an HNN involves initializing a neural network with random weights and employing a supermask to selectively retain critical weights. This results in a subnetwork, used during inference, which combines the initial network with the supermask, as shown in Fig. \ref{HNN_intro}. Despite using untrained weights, HNNs achieve accuracies on par with conventionally trained networks on complex datasets like ImageNet.

The random generation of weights in HNNs, achieved using simple hardware circuits, eliminates the need for energy-intensive data transfers from DRAM, allowing for 'almost-free' weight usage. Moreover, the supermask used in HNNs, being sparse and binary, can be conveniently stored on-chip. These attributes significantly alleviate the bottleneck typically associated with off-chip weight access and facilitate a more effective implementation of LPT, especially considering the reduced costs associated with weight reloading.


\section{Algorithm Realm: Layer-Penetrative Tiling}
While Hiddenite showcased the potential of HNNs, it was notably limited by its reliance on extensive on-chip SRAM storage for intermediate results, leading to high power consumption and area usage. Addressing these limitations, we propose the integration of Layer-Penetrative Tiling (LPT) as the processing flow, coupled with Blocked HNN with Tile Concatenation as the backbone structure. This combination effectively resolves the issue in past CL computations and block convolutions\cite{BlockConv}, resulting in a remarkable 14.2$\times$ reduction in the required on-chip activation memory capacity compared to Hiddenite. 

\subsection{Layer-Penetrative Tiling}
As illustrated in Fig. \ref{algorithm_concept}, we propose Layer Penetrative Tiling (LPT), an \textit{extreme} case of CL computation for fortified depth exploration, targeting significantly reduced storage requirement and energy consumption particularly for HNN. 
The LPT process starts with a single tile of input and penetrates through multiple layers (more than 10) in depth. LPT ensures that computation for the next tile commences only after the entire computation of the current tile is completed. 
Recall that traditional CL computation was constrained to just two to three layers due to the Weight Reloading Issue, since all weights spanning across layers needed to be stored on-chip for cost-effective inference. LPT, however, effectively overcomes this challenge. With our HNN-based accelerator design, weight data is generated directly on-chip, enabling almost 'free' transmission of these weights. This advancement paves the way for a more aggressive approach in deepening the integration of layers, which leads to a marked improvement in energy efficiency.

\subsection{Blocked HNN with Tile Concatenation}
Another challenge in conventional CL computation is the Data Dependency Issue, which arises because convolution operations require activation data from neighboring tiles, leading to a substantial amount of activation data being written to and retrieved from buffer memory, potentially undermining the efficiency gains achieved by LPT.
To address this, we introduce Blocked HNN with Tile Concatenation. Blocked HNN incorporates block convolution\cite{BlockConv} into HNN's CONV layers. As illustrated in Fig. \ref{blockconv_intro}, which divides the input activations into multiple small tiles, each separated by padding, to ensure that convolution within each tile is an isolated operation. This independence eliminates the need for data from neighboring tiles and confines the data dependency within the boundaries of each tile. 



One limitation of the conventional blocked convolution, however, is the restriction on the maximum fused depth imposed by the tile size. When encountering a pooling layer or a CONV layer with a stride greater than 1, the tile size reduces, which could lead to distorted computation results. For instance, applying a CONV3x3 on a 2x2 tile result in substantial accuracy degradation. In the strategy outlined in \cite{BlockConv}, small tile sizes are addressed by switching to standard convolution; however, this approach falls short in achieving the desired memory reduction in hardware.

To address this issue, we introduce the Tile Concatenation (TC) technique. This method is applied after processing up to a certain layer, denoted as layer K, and the output from layer K is temporarily stored in additional memory. Then, the computation of an adjacent tile is restarted from layer 1 and continues up to layer K. Once this is complete, the previously stored output tile is retrieved and concatenated with the current output, effectively doubling the tile size. This larger, concatenated tile is then used for further computations beginning from layer K+1.
TC thus provides a practical solution to the tile size limitation in LPT, allowing for accurate and efficient processing even in scenarios where layer characteristics would otherwise lead to reduced tile sizes and potential computation inaccuracies.

As a result, our implementation of LPT with blocked HNN enables the end-to-end inference of ResNet50@ImageNet with a mere 72KB of activation memory. This signifies a 14.2x decrease on activation storage, which is elaborated in the following section.

\begin{figure}[tbp]
\centerline{\includegraphics[width=8.5cm]{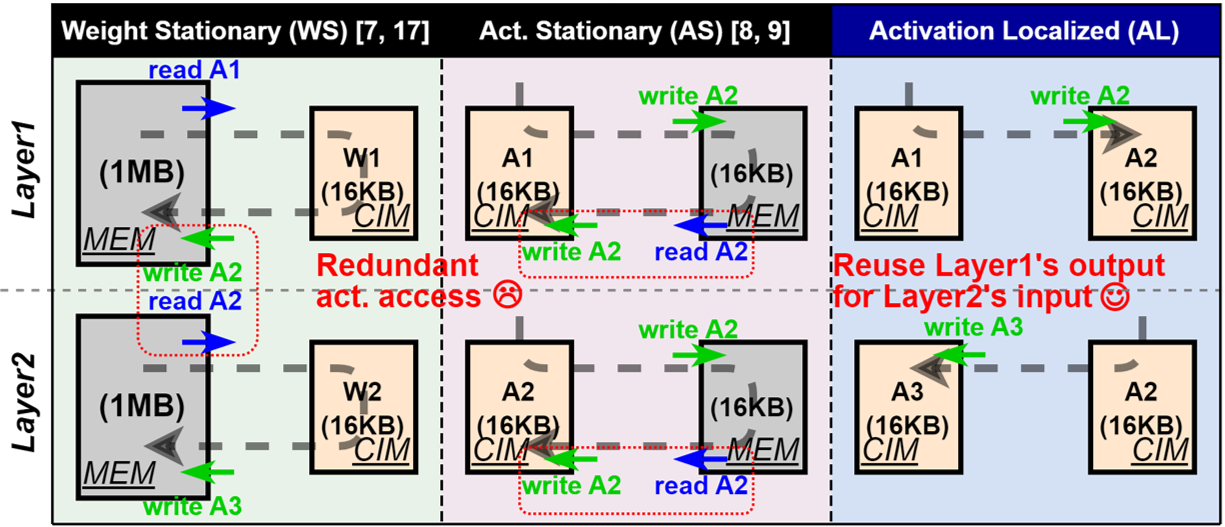}}
\captionsetup{font=footnotesize}
\caption{Both weight- and activation-stationary (WS and AS) dataflows incur redundant activation movement; Activation Localized (AL) dataflow avoids redundant activation accesses by localizing the intermediate results within CIM cores. 
}
\label{architecture_motivation}
\vspace{-4mm}
\end{figure}

\section{Architecture Realm: Activation-Localized CIM Processor}
Traditional NN accelerators, particularly those employing the CIM technique, often adopt a weight-stationary (WS) dataflow and incorporate large global activation memories to minimize substantial weight movement \cite{PIMCA, diana}. While such design effectively reduces weight movement, it leads to increased on-chip energy consumption due to the redundant activation movement \cite{DepFiN}.

\subsection{Activation-Localized Dataflow}
In response to this challenge, leveraging (1) the unique advantages of HNNs where the data movement of weights is almost 'free' and (2) the reduced storage requirement enabled by LPT, we shift our focus towards minimizing activation movement. To this end, we introduce an \textit{Activation-Localized (AL)} CIM dataflow, as shown in Fig. \ref{architecture_motivation}. This approach marks a distinct departure from the existing activation-stationary (AS) CIM structure \cite{ZPIM, GeneralPurposeCIM}, which only maintain activation stationary during computation. 
Our AL CIM dataflow extends this concept by preserving the locality of activations between the output of one layer and the input of the next. This is achieved by employing the CIM core as both the computation core and output memory interchangeably, allowing direct initiation of the next layer's computation with the output results stored in the CIM core. Such dual functionality of CIM core eliminates the need for additional transportation from the output memory back to the computation core, a requirement in the conventional AS dataflow, thereby effectively circumventing the dominant issue of communication power cost.


\begin{figure}[tbp]%
\centering
\subfigure[]{%
\label{ASCIM_architecture}%
\includegraphics[height=4.6cm]{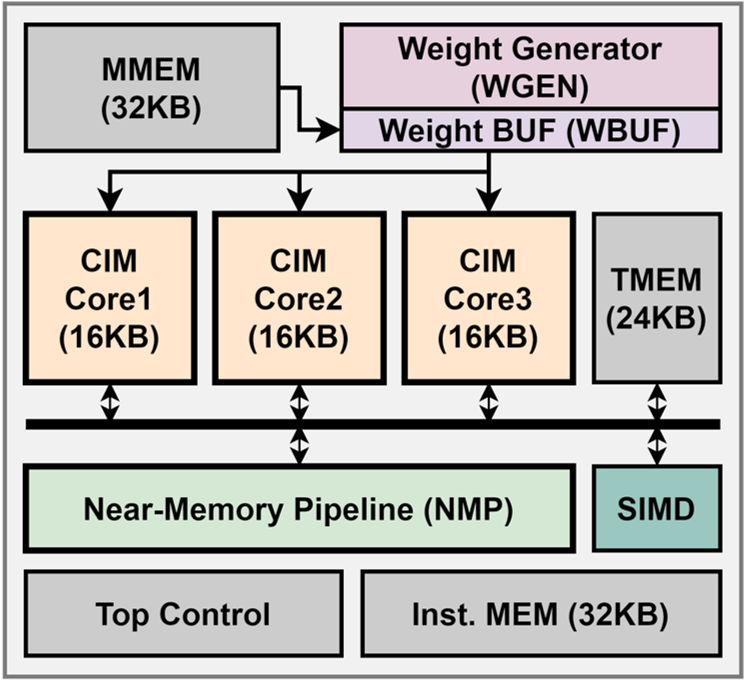}}
\subfigure[]{%
\label{AL_flow}%
\includegraphics[height=4.6cm]{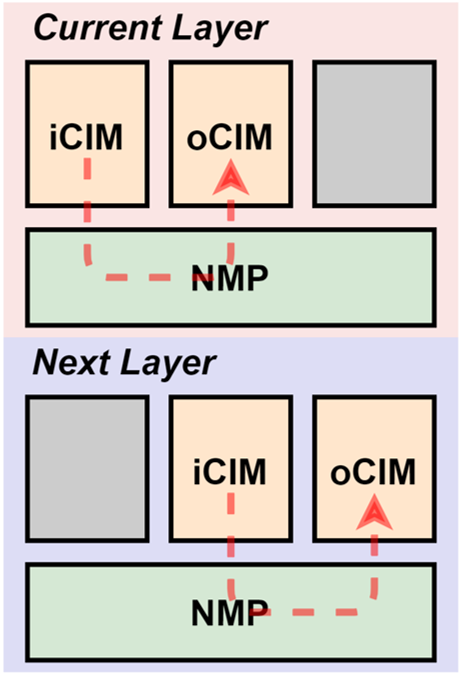}}%
\captionsetup{font=footnotesize}
\vspace{-4mm}
\caption{(a) The architecture of HALO-CAT, an Activation-Localized (AL) CIM HNN processor. (b) With AL dataflow, the cores act as iCIM and oCIM interchangeably to localize intermediate results.}
\vspace{-4mm}
\end{figure}

\subsection{Activation-Localized CIM HNN Processor}
The architecture of HALO-CAT, an activation-Localized (AL) CIM HNN processor, is depicted in Fig. \ref{ASCIM_architecture}, which features a meticulously designed for efficient HNN computation. The HALO-CAT key components and functionalities are as follows:\\
\textbf{CIM Cores:} Central to the processor are three 16KB CIM cores, each dedicated to AL computation. The cores are dynamically assigned roles; one serves as the input memory (iCIM) and another as the output memory (oCIM), facilitating a streamlined flow of activation data.
\\
\textbf{Near-Memory Pipeline (NMP):} Located proximately to the memory units, the NMP handles post-processing tasks. This strategic placement significantly reduces the distance data must travel, enhancing processing speed and energy efficiency.
\\
\textbf{HNN Weight Management:} The random weight generator (WGEN) is responsible for creating weights, which are subsequently masked in the weight buffer (WBUF). This masking is guided by the supermask data stored in the superMask memory (MMEM), ensuring precise weight application.
\\
\textbf{Operational Flow:} In operation, weights generated by WGEN are routed to iCIM for MAC operations. Post-MAC, the data undergoes further processing in the NMP before being transferred to oCIM for storage.
\\
\textbf{Layer Execution and Activation Localization:} To maintain activation locality, the roles of iCIM and oCIM are interchangeably assigned for each layer's execution, allowing seamless data transition between layers.
\\
\textbf{Support for Residual Connections:} The trio of CIM cores is specifically chosen to effectively support residual connections, a common feature in modern neural network architectures.
\\
\textbf{Tile Concatenation (TC) Mechanism:} To facilitate TC (as discussed in Section 3), TMEM, a 24KB Tile memory, temporarily stores intermediate tiles. These tiles are then concatenated within the CIM core, optimizing the computation process.
\\
\textbf{Memory Usage Efficiency:} Remarkably, the combined memory footprint, including the three CIM cores and TMEM, totals just 72KB, which is 14.2$\times$ less than Hiddenite. This demonstrates the processor's ability to manage complex computations within a compact and energy-efficient memory allocation.


\subsection{CIM Core Design}
\begin{figure}[tbp]
\centerline{\includegraphics[width=8.5cm]{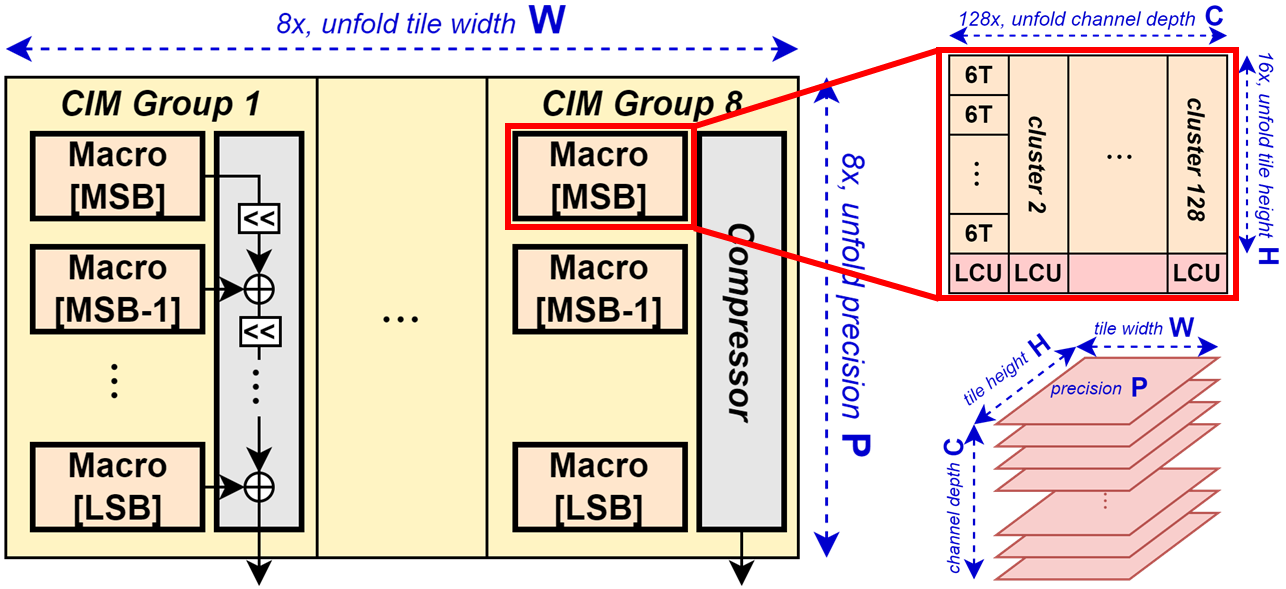}}
\captionsetup{font=footnotesize}
\caption{The structure and activation mapping of the CIM core.
}
\label{core_architecture}
\vspace{-4mm}
\end{figure}

The structure and activation mapping of HALO-CAT CIM core is illustrated in Fig. \ref{core_architecture}. The core is organized into eight columns of CIM groups, corresponding to the tile width $W$. Each group consists of a vertically aligned compressor and eight rows of macros, mapping to the activation precision $P$. These macros consist of 128 clusters and local computing units (LCUs), corresponding to the channel depth $C$, with each cluster containing 16 rows of 6T SRAM cells, unfolding to the tile height $H$. In addition, the core accommodates shallow channel depth by storing multiple pixels in one macro row and adjusts to deep channel depth by storing one pixel across several macro rows.

\begin{figure}[tbp]%
\centering
\subfigure[]{%
\label{nmc_architecture}%
\includegraphics[height=4.7cm]{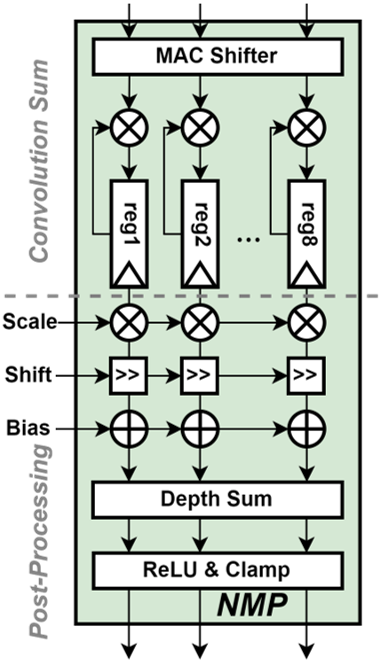}}
\subfigure[]{%
\label{conv_explanation}%
\includegraphics[height=4.7cm]{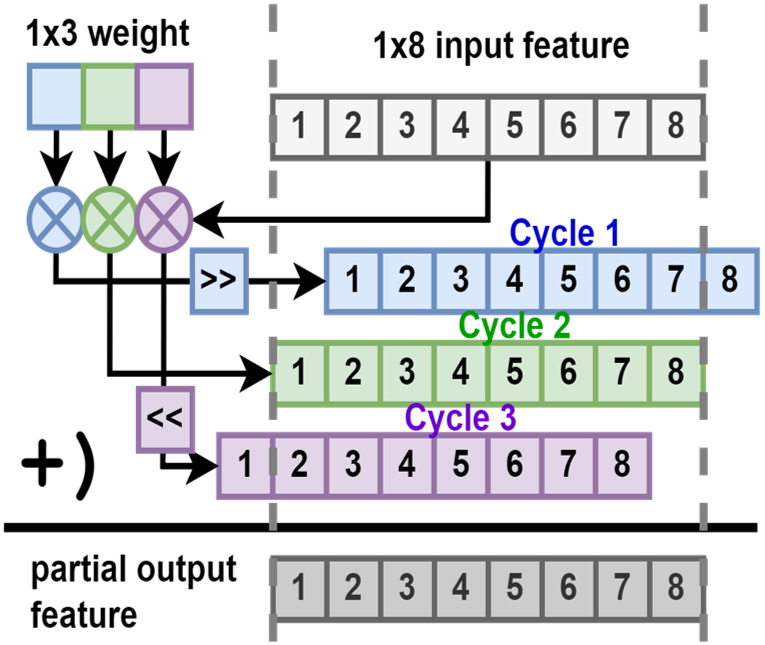}}%
\captionsetup{font=footnotesize}
\vspace{-4mm}
\caption{(a) The architecture of NMP. (b) NMP operation principles for CONV, which ensures optimal hardware utilization across diverse kernel sizes and obviates the need of extra input buffer.}
\vspace{-4mm}
\end{figure}

Every operation cycle activates one out of 16 SRAM rows, sending its data to the LCU for CIM computations across 128 SRAM columns. The resulting CIM outputs are subsequently distributed to the compressor and aggregated among 8 macros to generate MAC outputs for all 8 CIM groups in parallel. Lastly, these 8 channels of MAC outputs are transmitted to the NMP for convolution summation and post-processing. The execution of $H$ is bit-serial, while $C$, $P$, and $W$ are processed in a bit-parallel manner, optimizing computational efficiency.

Our AL CIM operational principle is ideally suited for CONV operations, eliminating unnecessary data accesses. For $K \times K$ CONV, the weight is partitioned into $K^2$ pixels, each sequentially processed in iCIM. This is illustrated in Fig. \ref{conv_explanation}, where a 1x3 weight interacts with a 1x8 input tile, producing an array of partial products (PPs). These PPs are conditionally shifted and accumulated in the NMP (Fig.\ref{nmc_architecture}), with MAC shifters facilitating efficient data manipulation. A standard CONV3x3 operation, for instance, involves repeating this process thrice with varied operand rows. Post-processing tasks like scaling, ReLU, and quantization are then executed in the NMP, before storing the results back to oCIM.

Contrary to most WS CIM architectures \cite{PIMCA, diana}, which require additional input buffers for activation reuse and often struggle to maintain high utilization rates for different kernel sizes, our dataflow ensures optimal utilization for any kernel size and negates the need for extra buffering.

\subsection{CIM Circuit Design}

\begin{figure}[tbp]%
\centering
\subfigure[]{%
\label{circuit_architecture}%
\includegraphics[height=3.6cm]{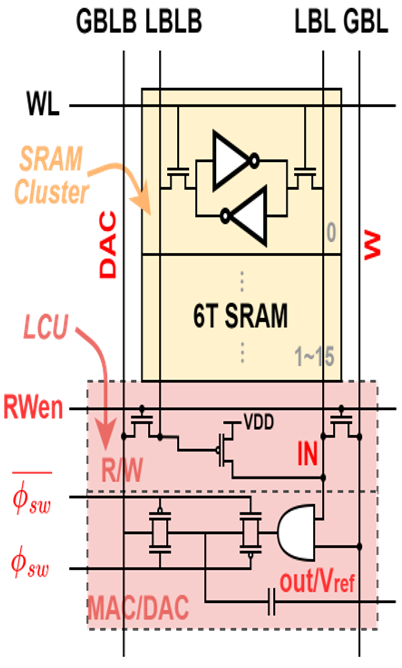}}
\subfigure[]{%
\label{resnet50_tile_config}%
\includegraphics[height=3.6cm]{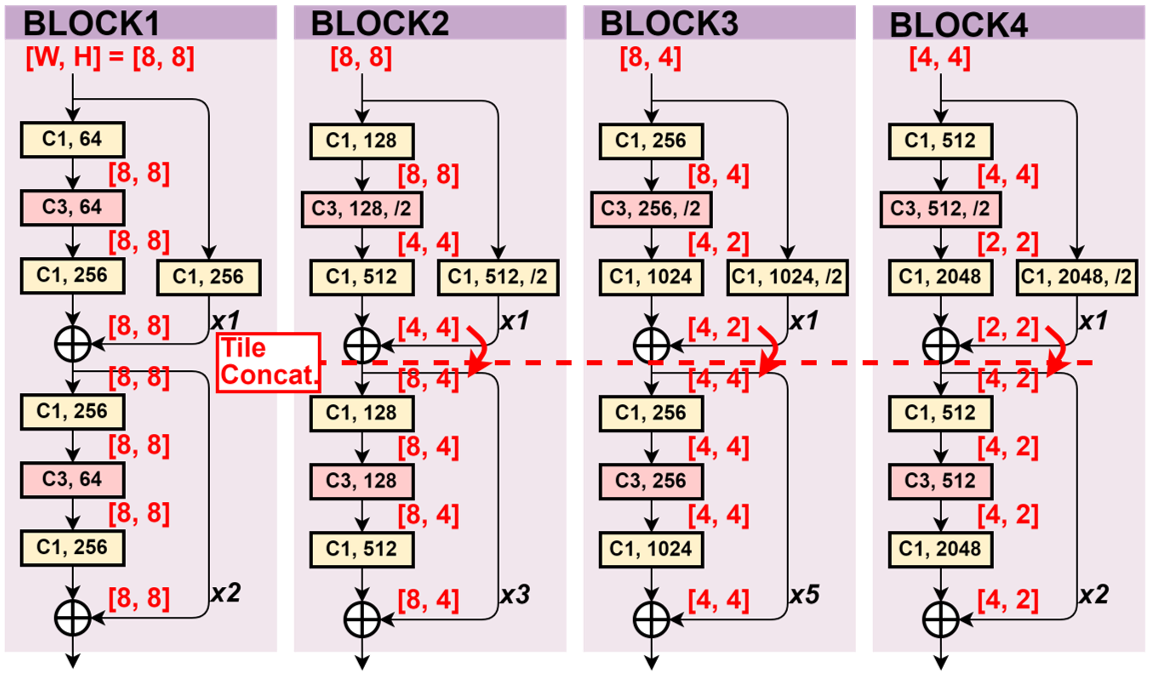}}%
\captionsetup{font=footnotesize}
\vspace{-4mm}
\caption{(a) Schematic of SRAM cluster and LCU. The capacitor is reused for both MAC computation and reference voltage generation, realized by two transmission gates with complementary control signals. (b) Tile sizes for LPT and utilization of TC under ResNet50@ImageNet.}
\vspace{-4mm}
\end{figure}

Fig. \ref{circuit_architecture} illustrates HALO-CAT's CIM circuit, comprising an SRAM Cluster with 16 bitcells, alongside a local computing unit (LCU). In pursuit of a more compact CIM core, we implement intra-macro A/D conversion method for our analog-domain CIM (ACIM). This approach negates the necessity for a large capacitor array in conventional SAR-ADCs. We repurpose the CIM compute capacitor for two functions: charge sharing and DAC reference voltage generation. During the CIM operation phase, activation is fetched using LBL, while weight and DAC signals are dispatched through GBL and GBLB. By selectively switching between $IN*W$ and DAC signal, we can efficiently derive both the MAC output and reference voltage using the same capacitors. While similar A/D conversion scheme has been employed for CIM with bitcell-wise computation in previous works \cite{CRCIM, CYYAO}, we are the first to integrate it into CIM with cluster-wise computation, resulting in an even more compact area density and efficient design.


%

\section{Evaluation}

\begin{figure}[tbp]%
\centering
\subfigure[]{%
\label{block_conv_eval}%
\includegraphics[height=3.7cm]{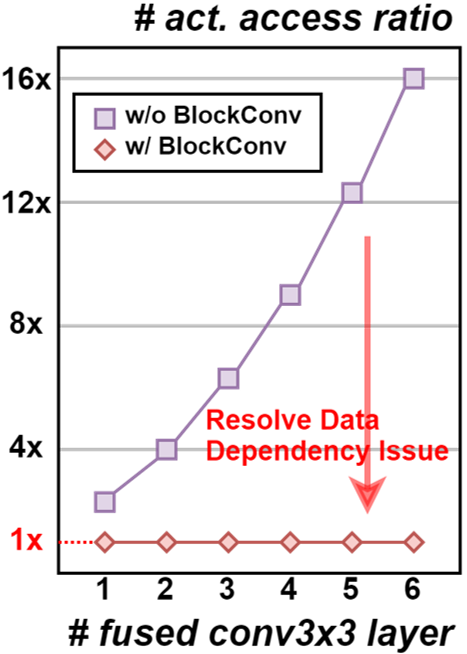}}
\
\subfigure[]{%
\label{layer_depth_eval}%
\includegraphics[height=3.7cm]{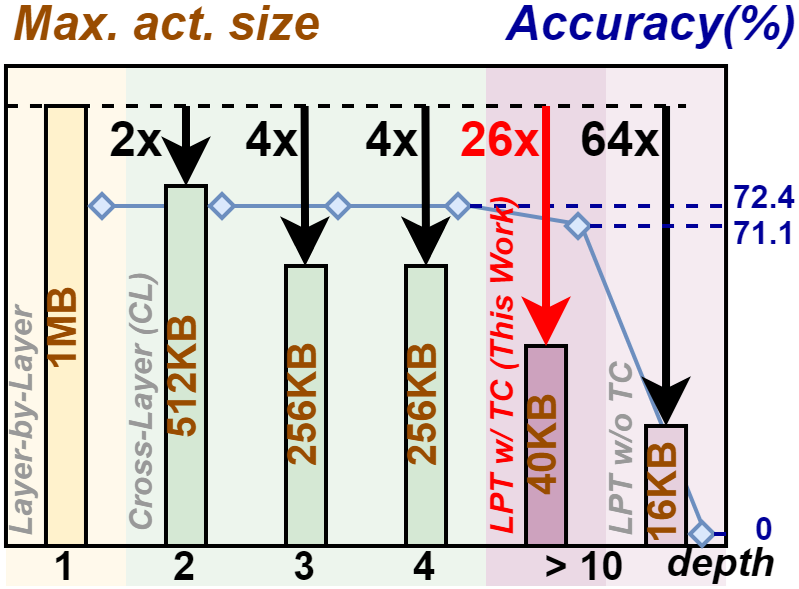}}%
\captionsetup{font=footnotesize}
\vspace{-4mm}
\caption{
(a) Simulation of \# activation access vs \# fused CONV3x3 layers, with and without block convolution for producing a 4x4 output tile. The use of block convolution achieves over 10$\times$ access reduction for deep fused-layers. 
(b) Max. activation size comparison of layer-by-layer, CL, and LPT processing flows. LPT, coupled with TC, achieves a 26$\times$ reduction in max. activation size while maintaining the accuracy loss within 2\%.
}
\vspace{-4mm}
\end{figure}

In the algorithmic domain, we assess how the integration of LPT and block convolution affects activation size reduction and activation access. Fig. \ref{block_conv_eval} demonstrates the impact on activation access across CONV3x3 layers for producing a 4x4 output tile, comparing scenarios with and without block convolutions. Without block convolutions, the number of activation accesses escalates significantly as the depth of fusion increases. In contrast, the incorporation of block convolution results in a consistent number of accesses even with deeply fused layers. This consistency results in a substantial reduction, by over 10 times, in activation accesses for deeply fused layers.
Further, Fig. \ref{layer_depth_eval} compares the maximum activation sizes required by various processing flows, including traditional layer-by-layer, CL, and our proposed LPT. While CL offers a 2 to 4$\times$ reduction, LPT achieves more substantial benefits, cutting down activation access by up to 26 to 64$\times$. However, LPT faces accuracy degradation with decreasing tile sizes, as described in Section 3.3. To address this, introducing TC in our approach effectively preserves ResNet50@Imagenet accuracy, with a drop of less than 2\% drop.

Furthermore, Fig. \ref{resnet50_tile_config} illustrates the specific tile configurations (width W and height H) applied in each layer for ResNet50@ImageNet inference. We resize the image length from 224 to 256 to simplify tiling efforts. TC is executed immediately after the first residual connection of blocks 2 to 4, a total of three times. This strategy, as opposed to performing TC right after strided operations, avoids concatenating results for both the main and the residual paths, leading to a 20\% reduction in the required size of Tiled Memory (TMEM).

\begin{figure}[tbp]%
\centering
\subfigure[]{%
\label{energy_table}%
\includegraphics[height=4.2cm]{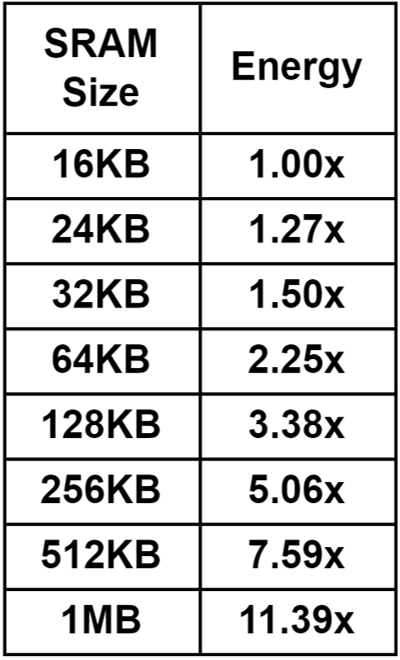}}
\
\subfigure[]{%
\label{dataflow_comparison}%
\includegraphics[height=4.2cm]{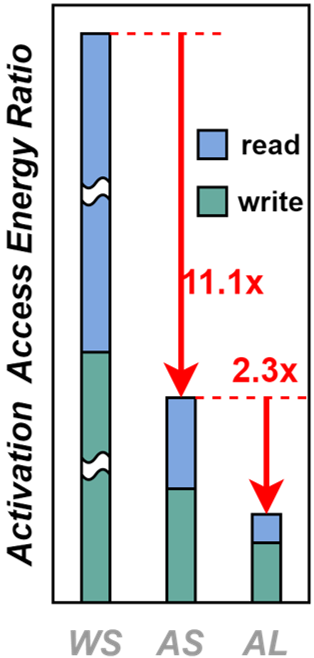}}%
\
\subfigure[]{%
\label{layout}%
\includegraphics[height=4cm]{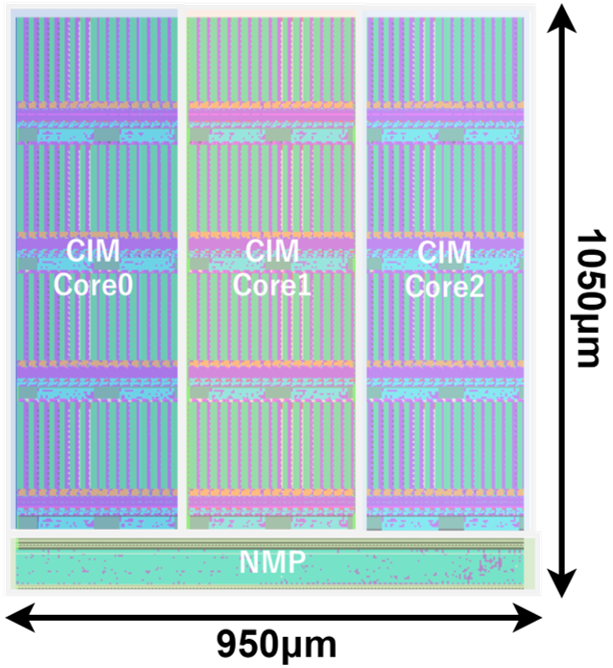}}%
\vspace{-2mm}
\subfigure[]{%
\label{baseline_comparison}%
\includegraphics[width=8.5cm]{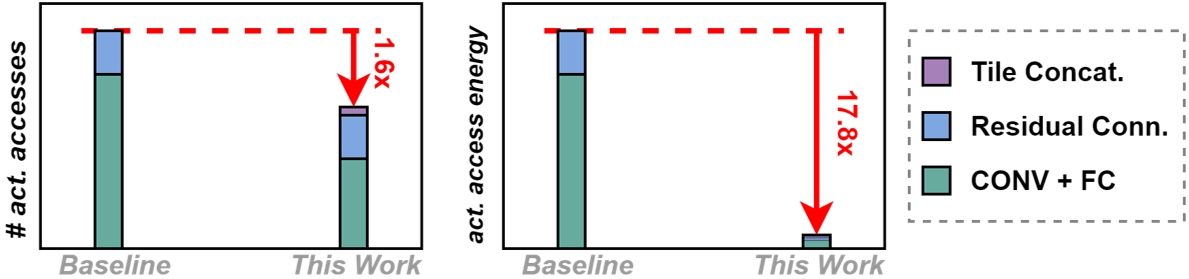}}%
\captionsetup{font=footnotesize}
\vspace{-4mm}
\caption{
(a) Estimated SRAM data access energy ratio
(b) Comparison of activation access energy between WS, AS, and AL dataflows. AL dataflow obtains a 2.3$\times$ reduction by preserving the activation locality between the output of the current layer and the input of the next.
(c) Layout of the computational components.
(d) Comparison between our work and the baseline implementation. HALO-CAT achieves a 1.6x reduction in \# act. accesses and a 17.8x reduction in act. access energy. In addition, the overhead incurred by TC is negligible.
}
\vspace{-4mm}
\end{figure}

In the architectural domain, our focus was on evaluating the energy associated with activation access across various dataflows.
To estimate the energy consumed by activation movement, we extrapolated the data from \cite{Interstellar} to cover a broader range of sizes, as depicted in Fig. \ref{energy_table}. Under our assumptions, standard SRAM read/write consume energy equivalent to the values listed in the table.
Fig. \ref{dataflow_comparison} presents a comparison of the estimated activation access energy for WS, AS, and our proposed AL dataflow. For WS, we assume a 1MB activation memory, commonly required for stabilizing the weights. On the other hand, AS and AL benefit from LPT, requiring only 16KB for tile storage. Our evaluation, specifically for end-to-end inference using AS, indicates an estimated 11.1$\times$ reduction in activation access energy compared to WS. Furthermore, our AL dataflow optimizes this further by eliminating the redundant activation accesses found in AS, contributing to an additional 2.3$\times$ energy improvement. 

To benchmark the system-level enhancement achieved by HALO-CAT, with a focus on optimizing activation access energy, we have established a baseline model for comparison, based on the Hiddenite framework \cite{Hiddenite}. Concerning off-chip communication, HALO-CAT requires the same number of external accesses with the baseline model. For the on-chip computation, the baseline model adopts slice-based layer fusion \cite{Hiddenite} and also prioritizes the activations' stationarity for a fair comparison. The evaluation considers an architecture comprising a single 1MB AMEM and a PE, focusing solely on the data access of AMEM. Fig. \ref{baseline_comparison} details our evaluation of activation access energy consumption for end-to-end inference, in line with the data presented in Fig. \ref{energy_table}. Notably, the implementation of the AL dataflow results in a 1.6x reduction in activation accesses. Furthermore, the memory requirement reduction achieved through LPT plays a significant role in energy conservation. Collectively, these enhancements lead to an overall reduction in energy consumption by a factor of 17.8x compared to the baseline model.

\begin{figure}[tbp]
\centerline{\includegraphics[width=8.5cm]{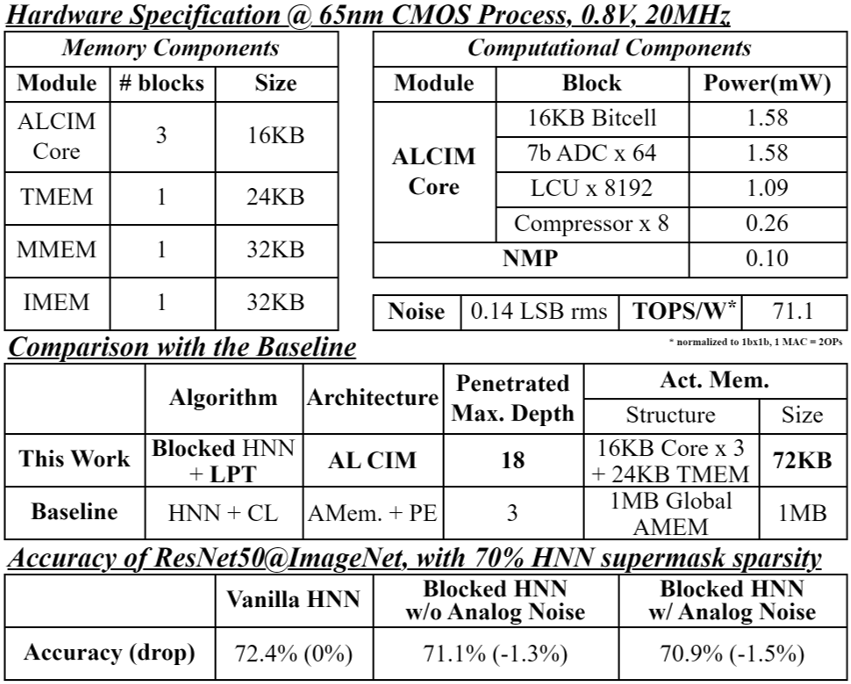}}
\captionsetup{font=footnotesize}
\caption{Summary and specifications of HALO-CAT. 
}
\label{summary_table}
\vspace{-4mm}
\end{figure}

Finally, we present the layout of the computational components, as depicted in Fig. \ref{layout}, and detail the circuit simulation results under 65nm CMOS process in Fig. \ref{summary_table}. 
Our architecture achieves 70.9\% accuracy on ResNet50@ImageNet, marking only a marginal  1.5\% decrease compared to the baseline HNN model. This demonstrates the robustness of our design while balancing computational efficiency and accuracy. In addition, the software-side development of HNN is still in its early stages, with ample room for improvement in areas such as supermask training and weight parameter generation\cite{okoshi2022multicoated}. We anticipate a significant enhancement in accuracy as advancements in these area continue to unfold.

\section{Conclusions}
In this paper, we introduced HALO-CAT, a hardware-software co-designed processor that synergistically integrates algorithmic and architectural advancements for HNNs. Algorithmically, we proposed LPT, an extreme case of CL tiling with aggressive depth exploration tailored for HNN. Coupled with block convolution and TC, LPT achieved a 14.2$\times$ reduction in activation memory capacity. Architecturally, we proposed the AL CIM processor to efficiently localize data between successive layers, featuring a 1.6$\times$ reduction in activation access. 
These innovations lead to a 17.8$\times$ reduction in estimated activation access energy, while the accuracy loss kept within 1.5\%, as evaluated on ResNet50@ImageNet.

\bibliographystyle{acm-num}
\bibliography{mybib}

\end{document}